\documentclass[twoside,twocolumn,9pt]{article}
\usepackage{extsizes}
\usepackage[super,sort&compress,comma]{natbib} 
\usepackage[version=3]{mhchem}
\usepackage[left=1.5cm, right=1.5cm, top=1.785cm, bottom=2.0cm]{geometry}
\usepackage{balance}
\usepackage{times,mathptmx}
\usepackage{sectsty}
\usepackage{graphicx} 
\usepackage{lastpage}
\usepackage[format=plain,justification=raggedright,singlelinecheck=false,font={stretch=1.125,small,sf},labelfont=bf,labelsep=space]{caption}
\usepackage{float}
\usepackage{fancyhdr}
\usepackage{fnpos}
\usepackage[english]{babel}
\usepackage{array}
\usepackage{droidsans}
\usepackage{charter}
\usepackage[T1]{fontenc}
\usepackage[usenames,dvipsnames]{xcolor}
\usepackage{setspace}

\usepackage{algorithm}
\usepackage{algpseudocode}
\usepackage{float}

\usepackage[compact]{titlesec}

\usepackage{epstopdf}

\definecolor{cream}{RGB}{222,217,201}

\begin{document}


\makeFNbottom
\makeatletter
\renewcommand\LARGE{\@setfontsize\LARGE{15pt}{17}}
\renewcommand\Large{\@setfontsize\Large{12pt}{14}}
\renewcommand\large{\@setfontsize\large{10pt}{12}}
\renewcommand\footnotesize{\@setfontsize\footnotesize{7pt}{10}}
\makeatother

\renewcommand{\thefootnote}{\fnsymbol{footnote}}
\renewcommand\footnoterule{\vspace*{1pt}%
\color{cream}\hrule width 3.5in height 0.4pt \color{black}\vspace*{5pt}} 
\setcounter{secnumdepth}{5}

\makeatletter 
\renewcommand\@biblabel[1]{#1}            
\renewcommand\@makefntext[1]%
{\noindent\makebox[0pt][r]{\@thefnmark\,}#1}
\makeatother 
\renewcommand{\figurename}{\small{Fig.}~}
\sectionfont{\sffamily\Large}
\subsectionfont{\normalsize}
\subsubsectionfont{\bf}
\setstretch{1.125} 
\setlength{\skip\footins}{0.8cm}
\setlength{\footnotesep}{0.25cm}
\setlength{\jot}{10pt}
\titlespacing*{\section}{0pt}{4pt}{4pt}
\titlespacing*{\subsection}{0pt}{15pt}{1pt}

\renewcommand{\headrulewidth}{0pt} 
\renewcommand{\footrulewidth}{0pt}
\setlength{\arrayrulewidth}{1pt}
\setlength{\columnsep}{6.5mm}
\setlength\bibsep{1pt}

\makeatletter 
\newlength{\figrulesep} 
\setlength{\figrulesep}{0.5\textfloatsep} 

\newcommand{\topfigrule}{\vspace*{-1pt}%
\noindent{\color{cream}\rule[-\figrulesep]{\columnwidth}{1.5pt}} }

\newcommand{\botfigrule}{\vspace*{-2pt}%
\noindent{\color{cream}\rule[\figrulesep]{\columnwidth}{1.5pt}} }

\newcommand{\dblfigrule}{\vspace*{-1pt}%
\noindent{\color{cream}\rule[-\figrulesep]{\textwidth}{1.5pt}} }

\makeatother

\twocolumn[
  \begin{@twocolumnfalse}
\vspace{3cm}
\sffamily
\begin{tabular}{m{4.5cm} p{13.5cm} }

&\noindent\LARGE{\textbf{Low-temperature chemistry using the R-matrix method}} \\
\vspace{0.3cm} & \vspace{0.3cm} \\

 & \noindent\large{Jonathan Tennyson,$^{\ast}$\textit{$^{a}$} Laura K. McKemmish,\textit{$^{a}$} and Tom Rivlin\textit{$^{a}$}} \\

& \noindent\normalsize{

   Techniques for producing cold and ultracold molecules are enabling the study 
   of chemical reactions and scattering at the quantum scattering limit, with 
   only a few partial waves contributing to the incident channel, leading 
   to the observation
   and even full control of state-to-state collisions in this regime.   
   A new R-matrix formalism is presented for tackling problems involving low- and ultra-low energy collisions. This general formalism is particularly appropriate for slow collisions occurring on potential energy surfaces with deep wells. 
   The many resonance states
   make such systems hard to treat theoretically but offer the best prospects for novel physics: resonances are already being widely used to  
   control diatomic systems and should provide the route to steering ultracold reactions.
   Our R-matrix-based formalism builds on the progress made in variational calculations
   of molecular spectra by using these methods to provide wavefunctions for the whole
   system at short internuclear distances, (a regime known as the inner region). These wavefunctions 
   are used to construct collision energy-dependent R-matrices which can then be propagated to give cross sections at each collision energy. The method is formulated for ultracold collision systems with differing numbers of atoms. 
 } \\
\end{tabular}

 \end{@twocolumnfalse} \vspace{0.6cm}
]

\renewcommand*\rmdefault{bch}\normalfont\upshape
\rmfamily
\section*{}
\vspace{-1cm}


\footnotetext{\textit{$^{a}$~Department of Physics and Astronomy, University College London, London WC1E 6BT, UK}}
\footnotetext{$^{\ast}$ E-mail: j.tennyson@ucl.ac.uk}






\section{Introduction}
To paraphrase the recent review by Stuhl {\it et al.}, \cite{14StHuYe}
a quiet revolution is occurring at the border between atomic physics
and experimental quantum chemistry.  
There has been a rapid development of techniques for producing cold and even
ultracold molecules through techniques such as photoassociation of ultracold alkali atoms, buffer-gas cooling, Stark deceleration, evaporative cooling \cite{12StHuYe} and laser cooling \cite{10ShBaDe,14ZhCoWa}. This progress is now enabling the experimental study of chemical reactions
and scattering at the quantum scattering limit with only a few partial
waves contributing to the incident channel (e.g. Quemener and Julienne
\cite{12QuJuxx}). Moreover, the ability to perform these
experiments with non-thermal distributions comprising specific states
enables the observation and even full control of state-to-state
collision rates in this regime. This is perhaps the most elementary
study possible of scattering and reaction dynamics.\cite{14StHuYe}
The trapping \cite{98WeDeGu} and subsequent study of chemical
reactions \cite{12SiJaTa} involving cold or ultracold chemically important
molecules, such as OH \cite{12StHuYe} and CaH, has opened a whole range of possibilities that
can be explored in chemical and quantum mechanical control and
exploitation.\cite{12JiYexx}
These experimental breakthroughs demand equally
transformative theoretical methods for treating ultra cold reactions;
these are, for many cases, still lacking.\cite{12QuJuxx}

One important feature of ultracold reactions is the pronounced
structures present in the cross sections due to temporary formation of
long-lived quasi-bound states of the compound system, known as
resonances.  Resonances are ubiquitous in ultracold collisions
\cite{10ChGrJu,13MaQuGo} and offer the best opportunity for quantum
control\cite{10OsNiWa} and steering: they are already used to steer the formation of ultracold diatomic molecules: see, for example, Malony {\it
  et al.}\cite{14MoGrJi} Furthermore, elastic and inelastic\cite{01Bohn}
cross sections can dramatically change near resonances,\cite{09HuBeGo}
which directly influences the effectiveness of sympathetic cooling and
trap losses. These resonances can be manipulated using magnetic and
electric fields.\cite{06KoGoJu} Studying the structures of resonances in
ultracold systems has yielded interesting physics, such as chaos in Dyspronium atoms,\cite{15MaFeKa,14Julienne} universal scaling laws/
characteristic behaviour\cite{d2004limits,10LoOtSe} and, when three or
more bodies are involved, Efimov
resonances.\cite{11FeZeBe,09WaEsxx,06DiEsxx,06KrMaWa,09KnFeMa,10LoOtSe,09StDiGr,10FeGrxx,11BeZeHu}
There are already a number of examples of novel many-body
state physics\cite{08BlDaZw} such as Bose-Einstein
condensates (BECs),\cite{03GrReJi,03JoBaAl}
 Efimov trimers,  as well as experiments investigating the crossover region
between the superfluidity of bosons in BECs and the Cooper pairing of fermions in
Bardeen-Cooper-Schrieffer (BCS) theory.\cite{04BaAlRi,04BoKhCu}

The resonance structure of systems which form over deep wells\cite{08PeGaCo} in their
potentials which support many bound states is likely to be
particularly rich and thus offer the greatest potential for transformative
science. These deep wells also offer the most opportunity for deviations from previously identified universal characteristics. 
Here we propose a formalism explicitly designed to study such systems.

From a theoretical perspective, gas-phase, quantum reactive scattering
at room/high temperature is well studied. Time-dependent methods have
proved to be particularly powerful for these problems \cite{03AlCl}.
However time-dependent methods struggle at ultra-low collision energies
because of the long collision times involved; they are particularly
poor at treating resonances.  There are time-independent methods
available which have been used to treat low-energy collisions. The
general physics can often be elucidated using simplified model
theories \citep{01Bohn}. 

More molecule-specific theories include, in
particular, procedures which use hyperspherical coordinates and basis
set methods.  \cite{87PaPa,89LaLe} These theories have been developed
and applied to low-energy collisions; see Honvault {\it et al}
\cite{11HoJoGo}, and Pradhan {\it et al.}  \cite{14PrBaKe}, for example.
These methods have been used successfully to treat a number of slow
atom-diatom collision problems and are the closest in spirit to what
is proposed here. However, the hyperspherical methods generally involve
transforming the problem into a series of adiabatic potentials for
which solutions are then found.  While this approach has proved
numerically successful, it is not physically motivated and ultimately
involves approximations concerning the couplings between the curves
which are hard to overcome.

The idea behind the new proposed R-matrix method for heavy particle
scattering is the division of space into two regions: an
(energy-independent) inner region where most of the physics takes
place, plus an
outer region where the interactions are simple. In this inner region,
solutions can be obtained by adapting standard bound-state programs.
The R-matrix is then constructed on the boundary between these
regions. Energy-dependent solutions to the scattering problem can then
be obtained rapidly by propagating the R-matrix. First principles,
or calculable, R-matrix methods have
proved outstandingly successful for studies of light particle collisions
\cite{Burke2011,jt474} and are being increasingly used in nuclear physics
\cite{10DeBa}. However, such methods
have yet to be systematically applied to
heavy particle collisions. R-matrix methods were extensively used to
study chemical reaction in the 1980s but, apart from
proof-of-principle studies \cite{85BoGe}, these calculations simply
used (outer-region) R-matrix propagation over the entire coordinate
range\cite{wl80}. The proposal here is fundamentally different and is much closer
in spirit to the methods successfully used by many groups to study
electron collisions.

Our R-matrix based formalism builds on the progress made in
variational calculations of molecular spectra which are now being used
to obtain solutions up to and beyond dissociation for strongly bound
systems such as water \cite{jt230,lg01,jt472,jt494} and H$_3^+$
\cite{jt100,jt132,btc94,jt358,scc10}. Both these
systems support about a thousand bound vibrational states
and many hundreds of thousands of bound rotation-vibration
states for which solutions are also being found \cite{jt64,jc13,jtpoz}.  
These variational calculations
provide wavefunctions for the whole system at short internuclear
distances. Indeed, resonances 
for water \cite{jt494,13SzCsxx} and H$_3^+$ \cite{jt443}
have already been studied using these approaches and a complex absorbing potential.

There are now a variety of variational nuclear motion methods and
related computer programs available for solving these problems. Here
we focus on the codes used within our group: specifically the new code
{\sc Duo}, designed for open-shell, coupled-state diatomic problems
\cite{jt609}, {\sc DVR3D} for three-atom problems \cite{jt338} and its
four-atom relative {\sc WAVR4} \cite{jt339}, as well as the general
polyatomic code {\sc TROVE} \cite{07YuThJe.method,15YaYuxx.method}.
Our group has significant experience with producing spectroscopic
accuracy potential energy surfaces that are generally assumed to be
essential for quantitative predictions of ultracold collision physics
\cite{jt236,jt375,jt519,jt526}. Hutson \cite{07Hutson} presents an
interesting counter-argument, demonstrating that if there are
significant couplings to inelastic channels, then the sensitivity of
the final cross-section to the details of the potential energy surface
is reduced as the peaks in the cross-section due to the poles produced
by the resonances are suppressed.

In this paper, we present our proposed methodology and illustrate it for
the case of atom-atom collisions, utilising the new {\sc Duo} program
\cite{jt609} to obtain the inner region wavefunctions. This simple
system allows a proof-of-principle demonstration of our proposed
methodology. Furthermore, the availability of relevant theoretical
\cite{jt458} and experimental \cite{14EdBa.Ar2} results will allow thorough
benchmarking of our methodology. In particular, we are interested in explaining
the surprisingly high measured cross section for the quenching of metastable,
excited argon atoms by ultracold argon \cite{14EdBa.Ar2}.

\section{Theoretical background}

Within the Born-Oppenheimer (BO) approximation, the solution of the
reactive scattering problem divides into two steps: construction of
the global potential energy surfaces using standard quantum chemistry
methodologies, and solution of the nuclear motion problem on these
surfaces to produce scattering cross-sections and other properties of
the reaction. We will assume here an appropriate potential energy surface is already available and focus on the second part of this problem. 
The desired
`solution' for scattering problems is the probability of different
processes at a given collision energy. Note that generally, the actual
wavefunctions solving the relevant time-independent Schr\"{o}dinger
equation are not necessary; instead, the observable information is
embedded in quantities like the phase shifts, scattering $S$ matrix,
the $K$ and $T$ matrices and the cross sections.

At large separation between the colliding species, the full scattering
problems can be represented in terms of partial waves. The
distinguishing characteristic of cold and ultracold scattering
problems is that only a small number of these partial wave components
are needed to obtain a very good approximation to the full answer.  At
short separation between the colliding species, a few partial waves
are no longer sufficient to describe the physics, particularly when
the two species interact strongly, i.e. collide over a potential with
a deep well. Instead of trying to use a large number of partial waves,
we propose using an approach which treats these two regions separately
using methods that are optimal for each region.
Specifically, we utilise the powerful variational nuclear motion
programs discussed earlier to find collision energy-independent solutions to the
inner region problem, $\psi_k$, using a single diagonalisation. These
energy-independent inner-region wavefunctions are used to construct
the so-called R-matrix at the boundary $r=a$ which is given
in standard formulations\cite{Burke2011} as
\begin{equation}
R_{i,j}(E,a) =
\frac{1}{2a} \sum_k \frac{\omega_{k,i} \omega_{k,j}}{E-E_k},
\label{e:ermat}
\end{equation}
where 
$i$ and $j$ are the asymptotic channels, and $k$ runs over the inner
region solutions, and the $\psi_k$ functions have energy $E_k$ and amplitude on
the R-matrix boundary $\underline{\omega}_k$.  The coordinate
$r$ is a radial coordinate which asymptotically goes to dissociation
products. Inner region solutions can be obtained explicitly in terms
of this coordinate by, for example, working in Jacobi coordinates, or the
amplitudes can be obtained by use of a projection operator on the boundary.

\begin{figure*}
\centering
\includegraphics[scale=0.8]{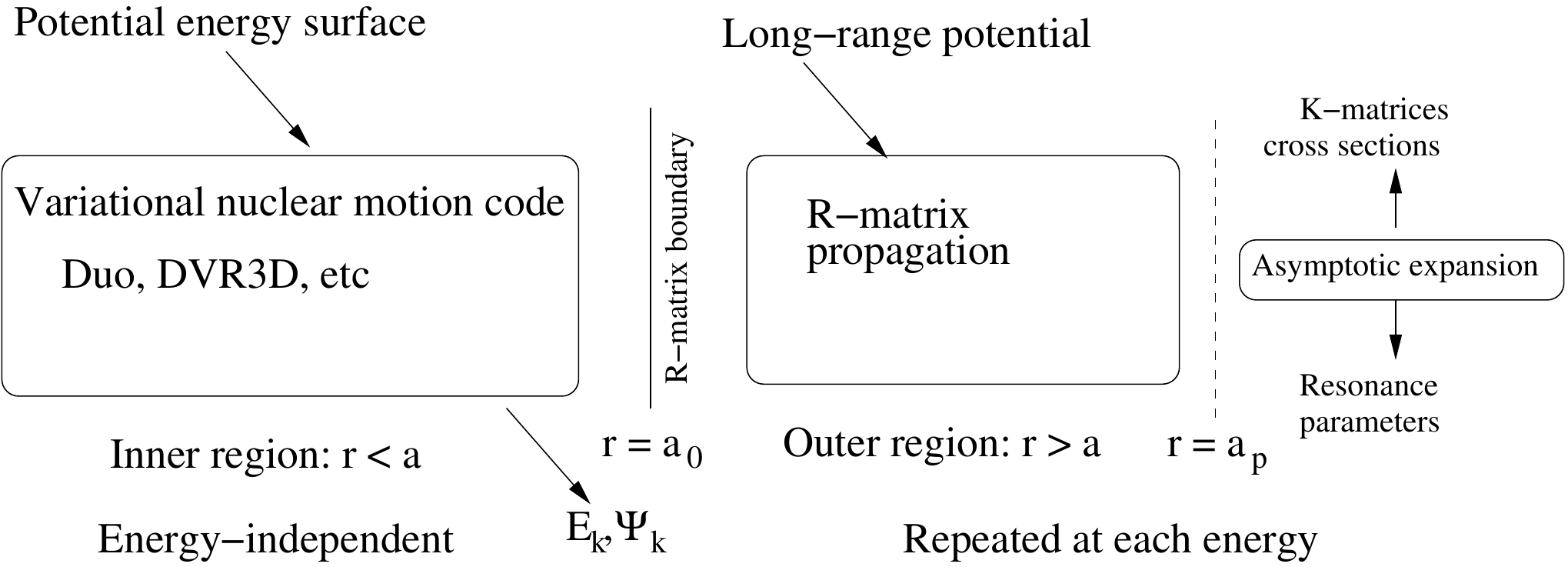}
 \caption
 {Schematic division of space illustrating the use of the R-matrix method.}  \label{f:flow}
\end{figure*}

Once the R-matrix has
been constructed at $r = a$, the energy-dependent, but computationally
simpler, outer region problem is solved to give K-matrices, from which
scattering observables, such as cross sections and resonance
parameters, can be determined. Due to the computational simplicity of this propagation,
this outer-region propagation can be performed on a fine grid of collision
energies, essential to elucidate resonance structure. Note that the
R-matrix propagation actually becomes simpler at colder
temperatures because the number of asymptotic channels decreases
significantly. Figure~\ref{f:flow} gives a schematic representation
of this solution strategy.

\section{Formal theory}

Below we develop the theory needed to solve a simple two atom collision
problem on a single potential energy curve. Such a theory might apply
to ultracold Ar -- Ar collisions. Note that while much of this
theory is standard, it is often given in atomic units (assuming
electron scattering)\cite{Burke2011,jt474}, such that the reduced mass terms, which are
important for heavy particle collisions, are missing.

Treating the inner region as a diatomic system, we can write a molecular ro-vibrational Hamiltonian operator in the following way:
\begin{equation}
\label{Hamilton}
\hat{H}^J = \frac{-\hbar^2}{2\mu}\frac{d^2}{dr^2} + \frac{\hbar^2J(J+1)}{2\mu r^2} + \sum_{i\geq i'} V_{ii'}(r),
\end{equation}
where $\mu$ is the reduced mass of the system of two masses $m_1$ and $m_2$:
\begin{equation}
\label{ReducedMass}
\mu = \frac{m_1m_2}{m_1 + m_2},
\end{equation}
$r$ is the internuclear separation, $J$ is the total angular momentum
of the molecule, and $V_{ii'}(r)$ is an element of the matrix of
potentials associated with the atomic channels, including the
off-diagonal channel coupling elements. These couplings can also
arise from effects such as spin-orbit interactions which can
be represented using a generalisation of rotational operator \cite{jt632}.
At this stage we are interested in both bound and continuum
solution to this problem.

Within the R-matrix method a (hyper-)radius $a_0$ is chosen where the
R-matrix is defined and inner regions solutions are obtained.
There is some flexibility over the choice of a$_0$, although our
plan is for the inner region to contain regions where the potential well
is deep. However we note that the R-matrix method has proved highly
successful at finding diffuse, long-range bound states which extend
outside the inner region\cite{jt560} and such states are expected
near the dissociation limit of polyatomic systems\cite{jt358}. 

Solving
the Schr\"odinger equation with the Hamiltonian defined in
eq.~(\ref{Hamilton}) within a finite region requires the introduction
of a surface term, $\mathcal{L}$, known as a Bloch
term\cite{blo57}, to retain Hermiticity. The expression for this term is:
\begin{equation}
\label{Bloch Equation}
\mathcal{L} = \frac{\hbar^2}{2\mu}\delta(r - a_0)\frac{d}{dr},
\end{equation}
where $\delta(r-a_0)$ is the standard Dirac delta function. To solve the molecular problem with the surface term, we introduce a set of functions $\{\chi_{n}^J(r)\}$. These functions are obtained as eigensolutions to the matrix problem
\begin{equation}
\label{ChiEquation}
(\chi_{n}^J | (\hat{H}^J + \mathcal{L}) | \chi_{n'}^J) = E_{n}^J \delta_{nn'}
\end{equation}
where, as is conventional\cite{Burke2011}, rounded Dirac brackets have
been used to show that integration in the radial coordinate, $r$, only
runs over the finite volume of the inner region, from $0$ to $a_0$. The 
eigenvalues, $E_n^J$, of this equation are usually referred to 
R-matrix poles and their associated eigenfunctions are defined using 
\begin{equation}
\label{ChiDefinition}
\chi_n^J(r) = \sum_i\sum_jc_{ijn}^J\phi_{ij}^J(r),
\end{equation}
where $\{\phi_{ij}^J(r)\}$ is some basis, and the coefficients
$c_{ijn}^J$ are determined by the requirement that
eq.~(\ref{ChiEquation}) is diagonal. Since the $J$ is a conserved
quantum number, we may label all solutions with it.  Final results require
the summation over $J$, but at low energies such sums should converge rapidly.

The indices $i$ and $j$ in eq.~\ref{ChiDefinition} run
over the channels, and the basis functions within each channel. 
To isolate the contribution from a single channel, one
can sum over only the basis functions within that channel, $j$, by
defining
\begin{equation}
\label{w_definition}
w_{in}(r) = \sum_{j}c_{ijn}\phi_{ij}(r).
\end{equation}
From this, elements of the R-matrix, $\mathbf{R}^J$, can be defined on the boundary using the heavy particle generalisation of eq.~(\ref{e:ermat})
\begin{equation}
\label{Rmat Equation}
R_{ii'}^J(E,a_0) = \frac{\hbar^2}{2 \mu a_0} \displaystyle\sum_{n} \frac{w_{in}^J(a_0)w_{i'n}^J(a_0)}{E_n^J - E}, 
\end{equation}
where $w_{in}^J(a_0)$ is called the \textit{surface amplitude} (since
it is evaluated at the boundary), $E$ is the scattering energy of
interest, and the sum is over all $n$, i.e. over all eigensolutions of
eq.~(\ref{ChiEquation}). We note that it is also possible to reformulate
the problem to use a reduced set of inner region solutions.\cite{jt332}
Note that a single set of inner region solutions
are used to construct the R-matrix at $r=a_0$ for {\it
  all} scattering energies, meaning that the inner region problem only
needs to be solved once, independent of how many energies the final
solutions are needed for. This is particularly useful for obtaining high-resolution
plots of resonances as a function of scattering energy.

From the scattering energy, $E$, the scattering wave number, $k$, can be defined as 
\begin{equation}
\label{k_definition}
k = \frac{\sqrt{2\mu E}}{\hbar}.
\end{equation}
A similar definition exists, and can be obtained from the eigenenergies $E_n^J$,
for the wave numbers $k_n^J$. These can be written into a diagonal matrix
$\mathbf{k}^J$.

Defining the outer region wavefunctions for a given target channel at some point, $r=a$, as $F_{i}^J(a)$, the R-matrix represents the relationship
between these functions and their derivatives:
\begin{equation}\label{eq:rmat}
F_{i}^J(a) = a \sum_{i'} R_{ii'}^J(E,a) \left.  \frac{dF_{i'}^J(r)}{dr} \right|_{r=a},
\end{equation}
where the sum runs over all channels. 

Propagating the R-matrix to large $r$ allows the scattering problem to
be solved without the explicit need to evaluate the wavefunction
which, particularly in the presence of closed channels, can be a difficult task numerically and computationally.

There are a number of means of propagating the R-matrix, including
those due to Baluja, Burke and Morgan \cite{bbm82,mor84}, due to Light
and Walker \cite{lw76}, and the software FARM (the flexible asymptotic R-matrix
package) \cite{farm,pfarm}. As discussed below, we favour the use of
the Light-Walker propagator. Furthermore, there are several ways of
obtaining the asymptotic wavefunctions, $F_{i}^J(r)$
\cite{bs62,gai76}. In this work the asymptotic expansion of Burke and
Schey \cite{bs62} is used. Generally speaking, asymptotic expansions
follow the form 
\begin{equation}
\label{AsymptoticExpansion}
F_i^J(r) = \displaystyle\sum_{i'}\left( s_{ii'}^J(k_{i'}^Jr) + \sum_{i''} c_{ii''}^J(k_{i'}^Jr) K_{i'i''}^J(E)\right),
\end{equation}
where both sums are over all channels, $K_{i'i''}^J$ is an element of
the K-matrix, $\mathbf{K}^J(E)$, and $s_{ii'}^J$ and $c_{ii''}^J$ are
elements of the matrices $\mathbf{s}^J(\mathbf{k}^Jr)$ and
$\mathbf{c}^J(\mathbf{k}^Jr)$ respectively. These matrices are matrices
of `sine-like' and `cosine-like' functions respectively, which are
different for different channels. The Burke-Schey asymptotic expansion
specifies the form of these functions, and is discussed in detail in
the next section.

The propagated R-matrix is then combined with the asymptotic expansion to construct the K-matrix, which has the following form:
\begin{equation}
\label{Kmatrix}
\mathbf{K}^J(E) = -\frac{\mathbf{s}^J(\mathbf{k}^Jr) - r \mathbf{R}^J(E,r) \mathbf{\dot{s}}^J(\mathbf{k}^Jr) }{\mathbf{c}^J(\mathbf{k}^Jr) 
- r \mathbf{R}^J(E,r) \mathbf{\dot{c}}^J(\mathbf{k}^Jr) },
\end{equation}
where $\mathbf{\dot{s}}^J(\mathbf{k}^Jr)$ and $\mathbf{\dot{c}}^J(\mathbf{k}^Jr)$ are the derivatives with respect to $r$ 
of $\mathbf{s}^J(\mathbf{k}^Jr)$ and $\mathbf{c}^J(\mathbf{k}^Jr)$ respectively, and $r$ is evaluated at some large value, denoted $a_p$.

From the K-matrix, the S- and T-matrices are defined in the following ways:
\begin{equation}
\label{S-Matrix_1}
\textbf{S}^J = \frac{\mathbf{1}+i\textbf{K}^J}{\mathbf{1}-i\textbf{K}^J},
\end{equation}
\begin{equation}
\label{T-Matrix_1}
\textbf{T}^J = \textbf{S}^J - \mathbf{1}.
\end{equation}
Note that while the definition of the S-matrix is general, the precise
definition of the T-matrix depends on the convention adopted.

The eigenphase for each channel, $\delta_i^J(E)$, is
given by the inverse tangent of $\mathbf{K}^J(E)$'s eigenvalues:
\begin{equation}
\label{lEigenphase}
\delta_i^J(E) = \arctan(K_i^J(E)),
\end{equation} 
where $K_i^J(E)$ is the $i^{\rm th}$ eigenvalue of the K-matrix, associated with channel $i$. This,
in turn, gives the eigenphase sum for a given symmetry ($J$):
\begin{equation}
\label{lEigenphasesum}
\delta^J(E) =  \sum_i\delta_i^J(E).
\end{equation}

The total cross section at a given energy, $\sigma_{\rm tot}(E)$, can
be obtained in a number of ways, including from the eigenphase sums:
\begin{equation}
\label{Total_Cross_Section}
\sigma_{\rm tot}(E) = \frac{4\pi}{k^2}\sum_{J=0}^{J_{\rm max}}(2J+1)\sin^2(\delta^J(E)),
\end{equation}
where $J_{\rm max}$ is the maximum number of angular momentum values
(partial waves) considered. For the ultracold temperatures being
considered here, this can be a very small number, possibly
a single channel.
For multi-channel collisions, the cross section for a
transition from channel $i$ to channel $i'$ is 
\begin{equation}
\label{Multichannel_Cross_Setion}
\sigma_{i'i}(E) = \frac{\pi}{k_i^2}\sum_{J=0}^{J_{\rm max}}(2J+1)|T_{i'i}^J|^2.
\end{equation}

\section{Computational Implementation}

The computational procedure for solving the above equations is
essentially made of three steps: (a) the inner region, (b) the
boundary and (c) the outer region and asymptopia. The final part, step
(c), can be written in a fairly general fashion, which should cater
for a variety of different systems. Therefore our aim in writing the
code which constructs the R-matrix on the boundary, step (b), is to
make it rather general to allow for the incorporation of a variety of
inner region nuclear motion codes. So far, in practice, we have only
used the diatomic code {\sc Duo} \cite{jt609}.  {\sc Duo} is designed
to compute spectra for open shell diatomic molecules and allows for
explicit inclusion of coupled potential energy curves through the
inclusion of spin-orbit and other coupling terms.  {\sc Duo} is
designed to read in potential energy and coupling curves in a variety
of formats, including simply as a grid of points. Here we have used the Ar --
Ar potential of Patkowski {\it et al} \cite{05PaMuFo.Ar2}, defined on
a grid.  {\sc Duo} constructs a basis using a Hund's case (a)
representation, which is then used to obtain a full variational
solution of the problem. Further details and discussion can be found
elsewhere \cite{jt589,jt632}.

The recently published version of {\sc Duo} \cite{jt609} is designed
only to treat bound rovibronic states. The first task is therefore to
extend this to give wavefunctions for the discretised continuum in the
inner region.  This is done by constructing a set of functions
$\{\phi_{ij}^J\}$. These functions are intended to be the set of square
integrable, linearly-independent basis functions which are complete
over the $[r_{min}, a_0]$ range up to some appropriate maximum energy
which enter into eq~(\ref{ChiDefinition}). As these functions
are used to provide the amplitude of the inner region function
$\{\chi_{n}^J\}$ on the boundary at $r=a_0$, one has to be careful how
these functions behave at this point.

\begin{figure*}[]
\includegraphics[scale = 0.4]{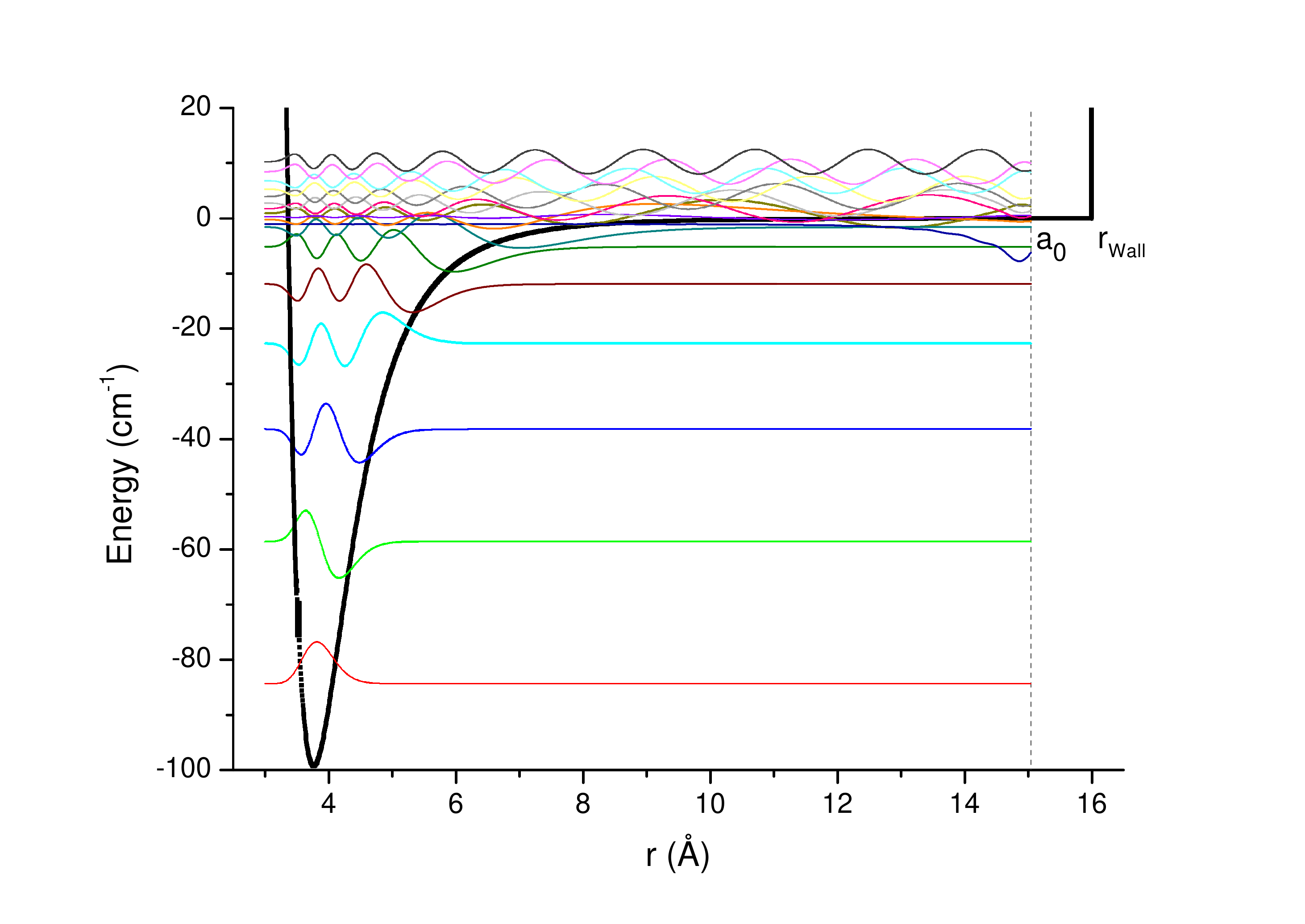}
 \caption
 {\label{fig:basisfunctions}Solutions of the inner region problem, $\chi_{n}^0$, for Ar$_2$
showing both bound and continuum functions.} \label{f:basis_functions}
\end{figure*}

In practice, the basis functions are generated in a two step procedure.
An initial basis set, $\{\psi_{ij}^J\}$, is generated by solving the
molecular problem associated with the ro-vibrational Hamiltonian,  $\hat{H}^J$, of
eq.~(\ref{Hamilton}). In solving this problem, an
artificial wall is placed in the potential at some distance $r_{\rm
  wall}$ ($ r_{\rm wall} > a_0$). Tests have shown that use of a wall
provides a good representation of resonance states contained inside it, see Fig.~\ref{fig:basisfunctions}.
Provided the wall is placed far enough out, the $\{\psi_{ij}^J\}$ are
effectively computational approximations of the eigenfunctions of
$\hat{H}^J$, with each basis function index $j$ belonging to a channel
$i$. Generally speaking, placing the wall such that $a_0$ was
approximately $\approx 95 \% $ of the way to $r_{\rm wall}$ was found
to be appropriate\cite{85BoGe}.

Although the $\{\psi_{ij}^J\}$ basis is constructed by integrating
over the full inner region $\left[r_{\rm min},r_{\rm wall}\right]$,
the rest of the R-matrix construction method involves integrating over
the smaller $\left[r_{\rm min},a_0\right]$ region. In this region, the
$\{\psi_{ij}^J\}$ are not eigenfunctions of the Hamiltonian, so
$\mathbf{H}^J$, the matrix of the Hamiltonian $\hat{H}^J$ in this
basis, will not be diagonal when defined over this range. The
non-diagonal Hamiltonian matrix is constructed by using a forward
finite difference numerical differentiation method of order four to
evaluate the kinetic term, and an implementation of Simpson's rule for
the numerical integration up to $a_0$.

The wall cannot be placed at
$a_0$ because, by construction, the basis functions have zero
amplitude at the wall. At this point there are two possible approaches,
both of which have been tested by us.

The earlier proof-of-principle study of an R-matrix approach to reaction
dynamics by Bocchetta and Gerratt\cite{85BoGe} simply diagonalised a generalised version of the eigenvalue eq.~(\ref{ChiEquation}) by
including the overlap matrix on the left-hand side. Alternatively,
the $\{\psi_{ij}^J\}$ functions can be
re-orthonormalised over the $\left[r_{\rm min},a_0\right]$ range, for
which we found symmetric or L\"owdin orthonormalisation\cite{50Lowdin}
to be the most suitable. Eventually we decided to utilise a generalised eigenvalue scheme, constructing the $\{\chi_{n}^J\}$ functions directly out of the $\{\psi_{ij}^J\}$. In both methods
we constructed a matrix for the Bloch operator in the original basis, 
$\mathcal{\mathbf{L}}$, using the aforementioned forwards finite
difference method of order four to compute the derivative at $a_0$.

The generalised eigenvalue problem we arrived at was
\begin{equation}
\label{GenEigVal}
(\mathbf{H}^J+\mathbf{L})\chi_{n}^J = E_n^J\mathbf{S}^J\chi_n^J,
\end{equation}
where $\mathbf{S}^J$ is the overlap, or Gramian matrix, whose elements are made
of all the possible inner products between the different $\psi_{ij}^J$ basis
functions. Equation (\ref{GenEigVal}) is then solved using the LAPACK\cite{99AnBaBi.method} routine
\textit{dsygv}  to obtain the $\{\chi_n^J\}$ and
$E_n^J$ required to construct the R-matrix.


For each value of $J$, our new R-matrix code reads from 
a version of {\sc Duo}, adapted to implement the potential wall, all of the
eigenenergies and eigenfunctions of the molecular system (the number of which is
user-specified in {\sc Duo}), the minimums $V_{ii\: \rm min}$ of the potentials
associated with each channel $V_{ii}(r)$, the
wall position $r_{\rm wall}$, the masses of the atoms $m_1$ and $m_2$, the
range over which the eigenfunctions are defined and orthonormalised,
$r_{\rm min}$ and $r_{\rm max}$, the step size of the integration, $\Delta r$,
and the zero point energy (zpe). 

This information is used to constuct the R-matrix on the boundary, as
outlined above, and this is then propagated outwards using the
Light-Walker propagation method to a point $a_p$. The Light-Walker
propagator takes the form of an iteration equation for the R-matrix
between the values $a_0$ and $a_p$, by dividing the region into
sub-regions with boundaries $a_s$. The propagator is constructed in
the following way\cite{Burke2011}: we diagonalise the
matrix
\begin{equation}
\label{Modified_Potential}
\mathbf{\mathcal{V}}^J(r) = \mathbf{V}^J(r) - \left(\mathbf{E}^J\right) + E\mathbf{I},
\end{equation}
where $\mathbf{I}$ is the identity matrix, $E$ is the scattering
energy, $\mathbf{E}^J$ is the diagonal matrix of eigenenergies (not to
be confused with the scattering energy), and $\mathbf{V}^J(r)$ is
the (in general) non-diagonal matrix of potentials for each channel,
including channel coupling elements (defined properly below in
eq.~(\ref{AsymptoticPotential}) -- note the $J$-dependence). We call
the version of this matrix which has been evaluated at $a_s$ and
diagonalised $\left(\mathbf{v}_s^J\right)^2$, and the matrix which
diagonalises it we call $\mathbf{O}_s^J$:
\begin{equation}
\label{Diagonalising}
\left(\mathbf{O}_s^J\right)^T\mathbf{\mathcal{V}}^J\mathbf{O}_s^J = \left(\mathbf{v}_s^J\right)^2.
\end{equation}

This allows us to define the real, diagonal matrix $\mathbf{\lambda}_{s}^J$ in the following way: 
\begin{equation}
\label{Lambda_definition}
\left(\mathbf{\lambda}_s^J\right)^2 = \frac{2\mu}{\hbar^2}\left( E\mathbf{I} - \left(\mathbf{v}_s^J\right)^2 \right).
\end{equation}

Next we define elements of the diagonal matrix $\mathbf{\mathcal{G}}_s^J$, made up of the following Green's functions:
\begin{equation}
\label{GreensFunction}
\begin{aligned}
\mathcal{G}_{is}^J(r,r') = \frac{-1}{\lambda_{is}^J\sin(\lambda_{is}^J\delta a_s)}  \\ \times \left\{ 
\begin{array}{cl}
\cos(\lambda_{is}^J(r'-a_s))\cos(\lambda_{is}^J(r - a_{s-1})) & a_{s-1} \leq r \leq r'\\
\cos(\lambda_{is}^J(r-a_s))\cos(\lambda_{is}^J(r' - a_{s-1})) & r' \leq r \leq a_s\\
\end{array} \right.,
\end{aligned}
\end{equation}
where $\delta a_s = a_s - a_{s-1}$. Then, defining $\mathbf{G}_s^J(r,r')$ as
\begin{equation}
\mathbf{G}_s^J(r,r') = \mathbf{O}_s^J\mathbf{\mathcal{G}}^J\left(\mathbf{O}_s^J\right)^T,
\end{equation}
we can write down the expression for the propagation equation: 
\begin{equation}
\label{Light-Walker}
\begin{aligned}
a_s\mathbf{R}_s^J = \mathbf{G}_s^J(a_{s},a_{s}) - \\ \mathbf{G}_s^J(a_s,a_{s-1}) \left( \mathbf{G}_s^J(a_{s-1},a_{s-1}) + a_{s-1}\mathbf{R}_{s-1}^J \right)^{-1} \mathbf{G}_s^J(a_{s-1},a_{s}),
\end{aligned}
\end{equation}
The size of each step in the iteration is variable, and dependent on
the size of the last step. It obeys its own iteration equation,
dependent on the derivative of the long-range potential used, in the
following way\cite{swl78}:
\begin{equation}
\label{stepsize}
\delta a_{s+1} = \beta \left( \frac{1}{N} \displaystyle\sum_{i=1}^{N}\frac{ \left(\lambda_{i,s}^J\right)^2 - \left(\lambda_{i,s-1}^J\right)^2}{\delta a_{s}} \right)^{-1/3},
\end{equation}
where $i$ counts over the channels, $N$ is the number of channels and, $\beta$ is a control parameter which allows you to specify how
many steps should be taken. $\beta$ is currently taken to be $0.1$.
The variable step size ensures that for different potentials, the
appropriate number of steps will be used to balance computation speed against accuracy. It also means that in the
multi-channel case, channels which contribute different amounts can be
treated differently.  The initial step size is taken to be $0.1\%$ of
the distance from $a_0$ to $a_p$.

Next we introduce the Burke-Schey expansion, a specifc version of
Eq.~(\ref{AsymptoticExpansion}). In the Burke-Schey expansion, the
matrices $\mathbf{s}^J$ and $\mathbf{c}^J$ have the following forms:

\begin{equation}
\label{BurkeExpansion}
s_{ii'}^J = A_{ii'}^J\cdot \sin(k_{i'}^Jr),
\quad 
c_{ii'}^J = B_{ii'}^J\cdot \cos(k_{i'}^Jr),\
\end{equation}
where 
\begin{equation}
\label{AlphaBetaSums}
A_{ii'}^J = \displaystyle\sum_{p=0}^{p_{\rm max}} \alpha_{pii'}^J r^{-p}, \quad B_{ii'}^J = \displaystyle\sum_{p=0}^{p_{\rm max}} \beta_{pii'}^J r^{-p},
\end{equation}
and the alpha and beta coefficients are derived from recurrence relations. For Ar$_2$, the test system being studied, the long-range Ar$_2$ potential is written as
\begin{equation}
\label{AsymptoticPotential}
V_{ii'}^J(r) = V_{ii'}(r) + \frac{\hbar^2J(J+1)}{2\mu r^2} = \displaystyle\sum_{\lambda=1}^{\lambda_{max}}a_{\lambda ii'} r^{-\lambda} + \frac{\hbar^2J(J+1)}{2\mu r^2},
\end{equation}
and the diagonal coefficients  $a_{\lambda ii}$ are obtained from Patkowski
and Murdachaew \cite{05PaMuFo.Ar2}. Note in general $\lambda_{\rm max}$ can vary
for different channels. The $\alpha_{pii'}^J$ and $\beta_{pii'}^J$ coefficients
are then obtained from the following interdependent recurrence relations :
\begin{equation}
\label{AlphaRecurrence}
\begin{aligned}
\left( \left(k_i^J\right)^2 - \left(k_{i'}^J\right)^2  \right)\alpha_{pii'}^J + ((p-1)(p-2) - J(J+1))\alpha_{p-2,ii'}^J \\
+ 2k_{i'}^J(p-1)\beta_{p-1,ii'}^J = \sum_{i''=1}^{N}\sum_{\lambda=1}^{\lambda_{\rm max}}a_{ii''\lambda}\alpha_{p-\lambda-1,i''i'}^J,
\end{aligned}
\end{equation}
and
\begin{equation}
\label{BetaRecurrence}
\begin{aligned}
\left( \left(k_i^J\right)^2 - \left(k_{i'}^J\right)^2  \right)\beta_{pii'}^J + ((p-1)(p-2) - J(J+1))\beta_{p-2,ii'}^J \\
- 2k_{i'}^J(p-1)\alpha_{p-1,ii'}^J = \sum_{i''=1}^{N}\sum_{\lambda=1}^{\lambda_{\rm max}}a_{ii''\lambda}^J\beta_{p-\lambda-1,i''i'}^J,
\end{aligned}.
\end{equation}
where $N$ is the number of channels and $\lambda_{\rm max}$ is the largest value
of $\lambda$ (with larger values increasing both accuracy and computation
time). The derivatives of the $\mathbf{s}^J$ and $\mathbf{c}^J$ matrices also
generate related recurrence relations, which can be derived by differentiating their
power expansions.

Finally, the coefficients obtained from the recurrence relations are used to
construct the asymptotic expansion. This expansion is combined with the R-matrix
to form the K-matrix using eq.~(\ref{Kmatrix}). From this K-matrix the
eigenphases are then obtained, and from the eigenphases the cross sections are obtained.

\section{Results}

As an initial test of the inner region codes we intend to use as
inputs to our new R-matrix code, we have looked for the so-called
shape resonances trapped behind the centrifugal barrier in the
rotationally excited Ar$_2$ problem. Tests were performed for the
rotational state $J=40$, which is sufficiently excited for the Ar$_2$
system to not support any truly bound states.  For this value
  of $J$, however, we would expect some quasibound states to exist
  behind a potential barrier, and for the Argon dimer potential of
Patkowski and Murdachaew\cite{05PaMuFo.Ar2}, the peak of the
centrifugal barrier is at $31.0389 \rm cm^{-1}$.  Calculations were
performed by inputting this Ar$_2$ potential on a grid both with Le
Roy's diatomic code {\sc LEVEL}\cite{lr07}
and with {\sc Duo}.

The {\sc Duo} results were obtained by computing bound and continuum
energy levels both with and without a wall using a stabilisation
procedure \cite{ht70}.  In the case of a wall energy levels were
obtained in {\sc Duo} up to $v=100$ at $J=40$ with the wall placed at
various locations.  A plot of various energy levels against wall
location was then constructed, and the places on that plot where the
energy levels appeared to lie on a horizontal line (i.e. where the
energy did not vary with wall location) were used determine the
energies of the shape resonances. This is because the continuum
energies follow a particle-in-a-box type energy distribution, and so
are dependent on the size of the `box'. This is not the case for
actual shape resonances.

A similar method was used to obtain the results without a wall, only instead of
plotting energy against wall position, a plot of energy against the position of
the end of the grid was used. Again, energy levels which did not vary with grid
size were taken to be resonances, as opposed to a continuum state. 

As Table \ref{t:Resonance_Test} shows, all three methods give two
resonance energy levels. Furthermore, all three
methods agreed to an accuracy better than $0.0001 \rm cm^{-1}$.  The
fact that in both {\sc Duo} cases the horizontally aligned energies
agreed with the LEVEL results was an encouraging indicator that
resonance energies had indeed been found.

\begin{table}[h!]
\centering
\caption{Ar$_2$ shape resonance energies obtained with three different inner region solution methods at $J=40$. 
All energies are in units of $\rm cm^{-1}$.} 
\label{t:Resonance_Test}
\begin{tabular}{llll}
\hline
  & {\sc LEVEL} & \multispan{2}{{\sc Duo}}\\\cline{2-4}
 N  &   &Wall& No wall\\
\hline
 1 & 7.7126       & 7.7126                 & 7.7126                    \\
 2 & 24.6178      & 24.6178                & 24.6178                 \\ \hline
\end{tabular}
\end{table}

Figure~\ref{fig:basisfunctions} shows a further test of the inner
region codes used as inputs for the R-matrix method, this time in the
form of Ar$_2$ inner region wavefunctions obtained for the inner
region problem using the Patkowski and Murdachaew potential curve in
{\sc Duo}. The wall was placed at 16~\AA\ and the inner region
boundary, a$_0$ at 15.045~\AA. We note that all bound state
wavefunctions, except the highest one, are completely confined well
inside our inner region boundary. The highest state's non-zero
  amplitude on the boundary is suspect and probably due  to residual
numerical issues. This means that the first 8 states
make no contribution to the R-matrix on the boundary and can be
dropped from consideration in the scattering region. For polyatomic
systems, dropping the truly bound states from consideration should
lead to substantial computational savings and could lead to the use
of methods which do not compute wavefunctions for these states in the
first place.

\section{Future Directions}

\begin{figure}[b!]
\centering
\includegraphics[scale=0.5]{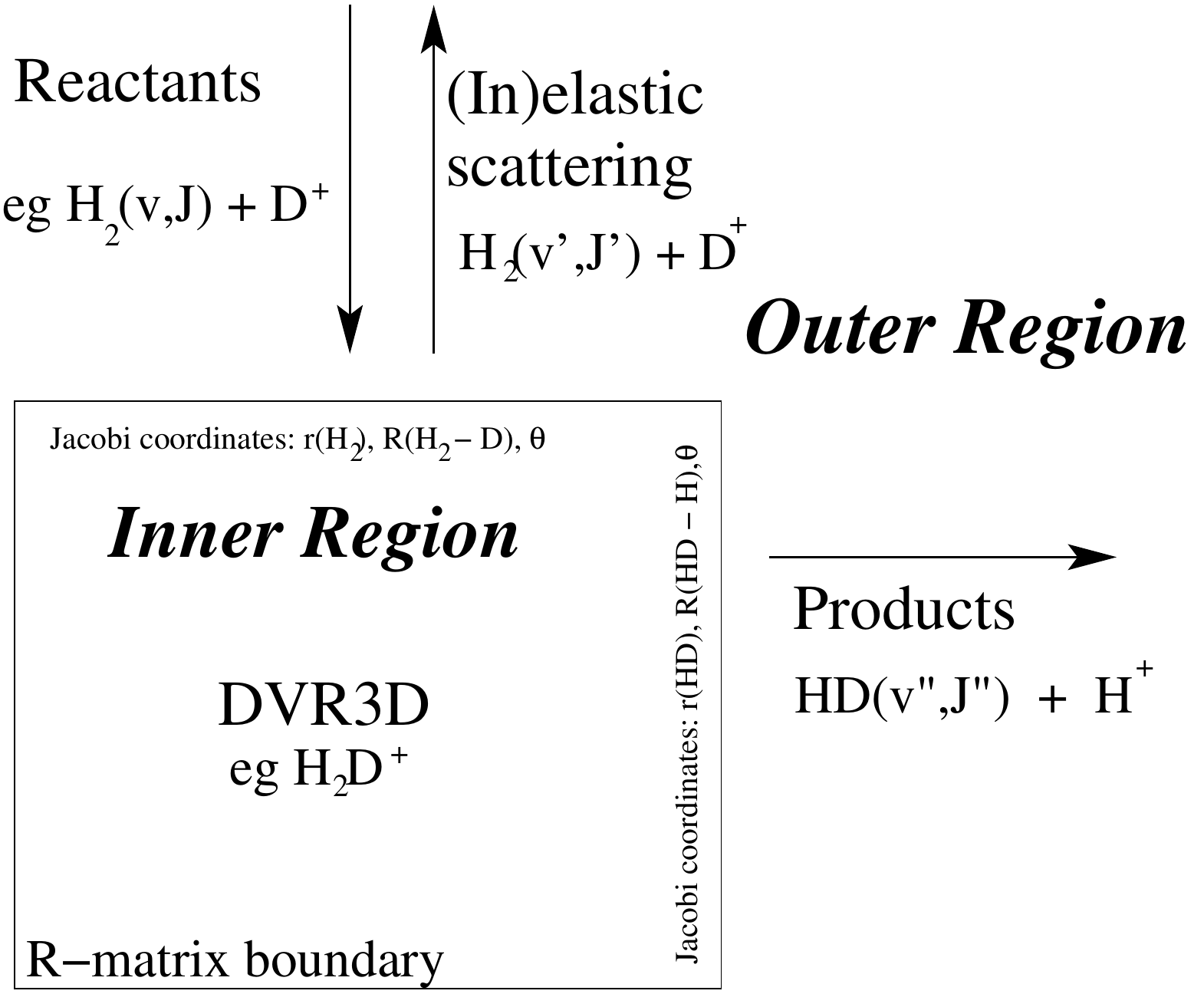}
 \caption
 {Schematic illustrating of the use of the R-matrix method.}  \label{f:rmat}
\end{figure}

Our aim is to use the R-matrix formulation of scattering theory to study
many-particle problems and, by extension, chemical reactions. As discussed
above, the methodology should be particularly appropriate for ultra-low energy
chemical reactions. Figure~\ref{f:rmat} illustrates how this should work
for the prototypical reaction
\begin{equation}
H_2 + D^+ \rightarrow HD + H^+.
\end{equation}
This exothermic reaction is likely to display significant resonance effects at very low
energies. We note that Fig.~\ref{f:rmat} implies there is a change
in coordinates, as the two asymptotes are most naturally represented in
different sets of Jacobi coordinates. How precisely this is best achieved
has to be determined, although one possibility would be to solve the inner
region problem in one coordinate system, here the higher symmetry 
H$_2$ -- D$^+$ Jacobi coordinates, for example, and then use a projection
operator on the boundary to construct the R-matrix on the boundary for
the other coordinates, in this case HD -- H$^+$.

In practice, our proposed methodology is by no means limited to
reactive scattering.  Table~\ref{processes} illustrates a number of the
possibilities, again using the H$_2$D$^+$ system as an example.

The division of space into two regions raises a number of interesting
possibilities. So far our studies on Ar -- Ar collisions have simply
used the same potential energy curves in the inner and outer regions.
However, the potential could be divided into two regions: an inner
region potential which captures the full complexity of the reaction, a
complex intermediate potential, and a long-range, outer region potential which
can be represented using known multipolar forms for the dissociation
fragment. Clearly, the two forms should match on the boundary. Standard
quantum chemistry methodologies can be used to produce these
potentials. 

\begin{table}
\caption{Processes that could be studied using a generalised R-matrix code: the H$_2$D$^+$ 
system is simply used as an example, and not all possible processes or products are listed.}
\begin{tabular}{ll}
\hline
Process & Example \\
\hline
Reactive scattering & D$^+$ + H$_2$ $\rightarrow$ HD + H$^+$\\
Photodissociation   &  H$_2$D$^+$ + $h\nu$ $\rightarrow$ HD + H$^+$ or  H$_2$ + D$^+$ \\
Photoassociation & H$_2$ + D$^+$  $\rightarrow$ H$_2$D$^+$ + $h\nu$\\
Charge Exchange & D$^+$ + H$_2$ $\rightarrow$ D + H$_2^+$\\
Elastic collisions &  D$^+$ + H$_2(v,J)$ $\rightarrow$ D$^+$ + H$_2(v,J)$ \\
Inelastic collisions &   D$^+$ + H$_2(v'',J'')$ $\rightarrow$ D$^+$ + H$_2(v',J')$\\
Predissociation & Not important for H$_2$D$^+$\\
\hline
\end{tabular}\label{processes}
\end{table}

Another possibility is to include very weak effects only in the outer
region.  For example, first principles studies of molecular spectra routinely
neglect hyperfine effects. However, these are important at ultra-low
collision energies.  In our proposed method one could re-couple
hyperfine-free inner region wavefunctions on the boundary so that the
outer region problem fully incorporates these effects. This approach
has been successfully used to treat spin-orbit effects in electron --
light atom collision problems for many years \cite{jajom}. A similar
approach could indeed be used to include spin-orbit effects, which are
usually totally quenched in strongly bound closed-shell systems, but
become important when dissociation occurs to open shell species, such
as what happens in water \cite{jt549}. Use of
the outer region in this fashion offers very significant
simplification compared to treating the full problem at all
internuclear separations.

Experimental study of ultracold molecules is significantly enhanced
through the use of electric and magnetic fields to tune resonances and
thus increase production rates of the ultracold molecules
(\cite{13QuBoxx}).
Similarly, weak field
effects could potentially be included in a similar fashion in the outer
region.  We note, however, that Zeeman effects have also recently
been included in {\sc Duo}.\cite{jtgfac}.

\section{Conclusions}

The possibility of studying cold and ultracold collisions processes,
and in particular chemical reactions, is one of the most interesting
developments of this century. These experiments are stimulating whole
new areas of scientific investigation, e.g. in quantum control, cold
collisions, cold chemistry, accurate measurement, tests of fundamental
physics, and more. Thus far most ultracold chemistry studies have been on alkali
metal dimers. Looking to the future, the next major stride will
involve reactions of chemically significant species and many atoms.
Particularly important for novel aspects of ultracold physics will be
the exploitation of resonances: long-lived quasi-bound states of the
compound system. The development of theoretical approaches, for
example the one described in this paper, are essential for predicting,
interpreting, and modelling
this new physics. The R-matrix approach is designed to
predict this interesting quantum behaviour and simulate and support
experimental studies in a rigorous and flexible manner, both theoretically
and computationally.

Our aim is to construct the harness code which links the inner and
outer region segments. Initially
this will be an atom-atom code used for testing numerical and
algorithmic aspects of the procedure; some of these results 
are presented here. This work
will be used to
guide the developments for larger collision systems. The atom-atom
code will also be used to study ultra-low energy collisions between
systems being studied experimentally, starting with the Ar -- Ar system
mentioned above; this will allow us to explore the treatment of
problems with coupled potentials and magnetic fields, and extend our work to other nuclear motion methods.


\section*{Acknowledgments}
This project has received funding from the European Union's Horizon
2020 research and innovation programme under the Marie
Sklodowska-Curie grant agreement No 701962 and from the EPSRC.

\bibliographystyle{rsc} 
\providecommand*{\mcitethebibliography}{\thebibliography}
\csname @ifundefined\endcsname{endmcitethebibliography}
{\let\endmcitethebibliography\endthebibliography}{}

\end{document}